\documentclass
[aps,twocolumn,prl,superscriptaddress,preprintnumbers,showpacs,tightenlines]{revtex4}%
\usepackage[dvips]{graphics, color}
\usepackage{amstext}
\usepackage{amsmath}
\usepackage{amssymb}
\usepackage{graphicx}
\usepackage{latexsym}
\usepackage{bm}
\usepackage{times}
\usepackage{amsfonts}%
\providecommand{\U}[1]{\protect\rule{.1in}{.1in}}
\definecolor{red}{rgb}{1,0,0}
\definecolor{blue}{rgb}{0,0,1}
\definecolor{bg}{rgb}{0,0,0.501}
\begin{document}
\title{Nondeterminstic ultrafast ground state cooling of a mechanical resonator}
\author{Yong Li}
\affiliation{Beijing Computational Science Research Center, Beijing 100084, China}
\author{Lian-Ao Wu}
\affiliation{Department of Theoretical Physics and History of Science, The Basque
Country University (EHU/UPV), P. O. Box 644, E-48080 Bilbao, Spain }
\affiliation{IKERBASQUE, Basque Foundation for Science, E-48011 Bilbao, Spain}
\author{Ying-Dan Wang}
\affiliation{Department of Physics, University of Basel, Klingelbergstrasse 82, 4056
Basel, Switzerland} \affiliation{Department of Physics, McGill University, Montreal
QC, H3A 2T8, Canada}
\author{Li-Ping Yang}
\affiliation{Institute of Theoretical Physics, Chinese Academy of Sciences, Beijing 100190, China}
\affiliation{Beijing Computational Science Research Center, Beijing 100084, China}

\begin{abstract}
We present an ultrafast feasible scheme for ground state cooling of a
mechanical resonator via repeated random time-interval measurements on an
auxiliary flux qubit. We find that the ground state cooling can be achieved
with \emph{several} such measurements. The cooling efficiency hardly depends
on the time-intervals between any two consecutive measurements. The scheme is
also robust against environmental noises.

\end{abstract}

\date{\today}
\maketitle


\emph{Introduction.}-- In the quantum regime, ground state cooling of a small
thermal object is an intriguing challenge and one of the most desirable
quantum technologies. Physically, the cooling process can be formulated as a
\emph{transformation} from a thermal state of the small object into its ground
state. The transformation is irreversible and cannot be performed when the
object is isolated.

A mechanical resonator~(MR) is a small mecroscopic mechanical object, usually
coupled to an auxiliary setup. The physical realization of its quantum ground
state has become more and more important in ultrahigh-precision measurements,
classical to quantum transitions, preparations of non-classical states,
quantum information processing \cite{ultrahigh
detection,quantum-mechanical,QIP-MR}. Throughout the years, considerable
number of optomechanical~\cite{Metzger:2004,Thompson:2007,Wilson-Rae:2007,
Marquardt:2007, Li-Cooling,Kippenberg2007,Li-2010} and
electromechanical~\cite{electromechanical-1,Zhang-PRL,electromechanical-3,electromechanical-4,Jacobs-2010,cooling-in-TLR}
models has been proposed for achieving their ground state cooling. Examples
are a bang-bang cooling through a Cooper pair box \cite{Zhang-PRL}, a
single-shot state-swapping cooling via the superconductor via a superconductor
\cite{Jacobs-2010}. Recently, some of us~\cite{Li-2010} proposed a ground
state cooling scheme of MR in an optomechanical system by controlling fast the
optical drives. The best studied ground state cooling protocol is the sideband
cooling~\cite{Metzger:2004,Wilson-Rae:2007, Marquardt:2007,
Li-Cooling,Kippenberg2007} designed originally for cooling the atomic spatial
motion. This cooling is now widely used for the MR cooling experiments and the
recent record for the mean phonon number is $3.8$~\cite{cooling-in-TLR}
Theoretically, a MR could be cooled down to its ground state in the resolved
sideband limit, with the mean phonon number less than 1. Since ground state
cooling is not yet achieved experimentally and seems to become much harder
when close to ground state, new cooling approaches remain desired.

This work presents a new protocol for MR cooling using repeated projective
measurements on an auxiliary qubit. While it may be used in any stage of the
cooling process of a MR, our protocol is aimed at the ground state cooling.
Our study starts with the cooling with repeated equal time-interval
measurements on the qubit, introduced for purification of quantum
states~\cite{purification} or measurement-based entanglement
generation~\cite{Wu-PRA04,Plenio-PRA99}. While equal time-interval repeated
measurements work well for ground state cooling, we find, unexpectedly, that
the cooling efficiency is even much better when the repeated measurements are
taken randomly. The ground state can be reached in several such measurements
in very short time. We give explanation on the unexpected phenomenon. These
results suggest that our protocol is completely robust against measurement
operational errors. In addition to this great advantage, the scheme is also
robust against environmental noises.

\begin{figure}[pt]
\begin{center}
\includegraphics[bb=80 136 460 400, scale=0.4,clip]{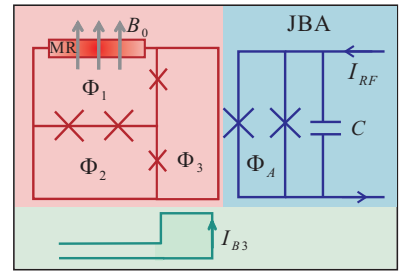}
\end{center}
\caption{(Color online) Schematic diagram of the circuit. The top left part (red) is
the coupled flux qubit-MR system, where each cross denotes a Josephson junction and
the bar denotes a doubly-clamped MR. The interaction between MR and flux qubit is
modulated by an in-plane magnetic field $B_{0}$. The top right part (blue) is a
Josephson bifurcation amplifier (JBA) formed by a dc SQUID shunted by a capacitance.
The bottom part (green) is bias to control the qubit energy gap as well as the
coupling
strength between the JBA and the qubit. }%
\label{fig:circuit}%
\end{figure}

\emph{Model.}-- We employ a gradiometer-type flux qubit~\cite{Paauw2009,Fedorov2010}
as our auxiliary qubit, though our scheme may be applicable to any two-level system
coupled to a MR. Fig.~\ref{fig:circuit} is a schematic diagram of our cooling setup.
The doubly-clamped MR (the red bar in Fig.~\ref{fig:circuit}), is embedded in a flux
qubit circuit which composed of three superconducting loops
with four Josephson junctions (JJs). An in-plane magnetic field $B_{0}$ induces
qubit-MR coupling via Lorentz force~\cite{Fei Xue}. The top right blue (bottom green)
part is to measure (operate) the qubit as described later.

The Hamiltonian of the flux qubit can be written as
$H_{q}={\hbar\Delta}\sigma_{x}/{2}+{\hbar\epsilon}\sigma_{z}/{2}$, where
$\sigma_{z}$ and $\sigma_{x}$ are the Pauli matrices in the basis of two
persistent current states $\left\vert \uparrow\right\rangle $ and $\left\vert
\downarrow\right\rangle $. Here, $\hbar\Delta$ is the tunneling amplitude
between the two states. The bias energy $\hbar\epsilon$ linearly depends on
external flux bias and in our case is set to zero by pre-trapping one flux
quantum $\Phi_{0}$ in the loop~\cite{Paauw2009,Fedorov2010}.

The MR is modeled as a single-mode harmonic oscillator with a high-$Q$ mode of
frequency $\omega_{m}$ and effective mass $m$. The entire system is characterized by
the Hamiltonian~\cite{Fei Xue}
\begin{equation}
H=H_{q}+\hbar\omega_{m}a^{\dagger}a-\hbar g(a+a^{\dagger})\sigma
_{z},\label{eq:1}%
\end{equation}
where $a$ and $a^{\dagger}$ are annihilation and creation operators for the MR with
frequency $\omega_{m}$. The last term denotes the interaction between the MR and the
flux qubit. The coupling constant is $g=B_{0}I_{p}L_{0}$ with $B_{0}$ the magnitude
of the in-plane magnetic field, $I_{p}$ the magnitude of the persistent current in
the loop, and $L_{0}$ the length of the MR.


Here we consider a MR with fundamental mode frequency $\omega_{m}\sim
2\pi\times100$~MHz. The qubit is tuned into resonance or near resonance with
the MR by monitoring the tunneling $\Delta$~\cite{Fedorov2010,Zhu2010}. The
coupling constant $g$, e.g., $\sim2\pi\times1$~MHz, is much smaller than the
qubit frequency such that
the rotating wave approximation~(RWA) can be used to reduce the Hamiltonian
Eq.~(\ref{eq:1}) to the standard Jaynes-Cummings (JC) Hamiltonian
\begin{equation}
H=\hbar\omega_{m}a^{\dagger}a+\frac{\hbar\Delta}{2}\tilde{\sigma}_{z}+\hbar
g(a\tilde{\sigma}_{+}+a^{\dagger}\tilde{\sigma}_{-}),\label{eq:3}%
\end{equation}
where $\tilde{\sigma}_{z,\pm}$ are the Pauli operators in the new basis of
ground and excited states: $\left\vert g/e\right\rangle =(\left\vert
\downarrow\right\rangle +/- \left\vert \uparrow\right\rangle )/\sqrt{2}$.

The operator $\hat{N}_{c}=a^{\dagger}a+\left\vert e\right\rangle \left\langle
e\right\vert $ in the JC model is conserved, such that $H$ can be represented
by the direct sum of a one-dimensional block in the basis $\left\vert
0,g\right\rangle $ and two-dimensional submatrices in pairs of bases
$\left\vert n,g\right\rangle $ and $\left\vert n-1,e\right\rangle $ when
$n\geq1$. The Hamiltonian (\ref{eq:3}) can be therefore diagonalized (we set
$\hbar=1$)
\begin{equation}
H=-\frac{\Delta}{2}\left\vert 0,g\right\rangle \left\langle 0,g\right\vert
+\sum_{n\geq1}\sum_{s=\pm}\varepsilon_{n}^{s}\left\vert ns\right\rangle
\left\langle ns\right\vert ,
\end{equation}
%
where the dressed eigenstates
$\left\vert n+\right\rangle =\cos\theta_{n}\left\vert n-1,e\right\rangle
+\sin\theta_{n}\left\vert n,g\right\rangle ,$ $\left\vert n-\right\rangle
=\sin\theta_{n}\left\vert n-1,e\right\rangle -\cos\theta_{n}\left\vert
n,g\right\rangle $ and the corresponding eigenvalues are $\varepsilon_{n}%
^{\pm}=(n-1/2)\omega_{m}\pm\sqrt{(\Delta-\omega_{m})^{2}/4+g^{2}n}$.
Here $\theta_{n}$ satisfies the equation $\tan2\theta_{n}=2g\sqrt{n}%
/(\Delta-\omega_{m})$ for $n\geq1$.

\emph{Unitary evolution and repeated measurements.}-- We prepare the whole
system initially in a separable state, $\rho_{0}=\left\vert g\right\rangle
\left\langle g\right\vert \otimes\rho_{m}$. Here
$\rho_{m}$ is the thermal state of the MR (it could also be an arbitrary
state, where the MR ground state $\left\vert 0\right\rangle $ is included). We
then perform repeated but unequal time-interval measurements (UTIMs) on the
flux qubit. The whole system evolves under the JC Hamiltonian~(\ref{eq:3}) in
between the measurements. The $j$-th measurement takes place at the time
instant $t_{j}=j\tau+\delta t_{j}$, where $\tau$ is a given time interval and
$\delta t_{j}$'s are random variations in time in the interval $(-\tau
/2,\tau/2)$. The time interval between the $(j-1)$-th and $j$-th measurements
is unequal, $\tau_{j}=\tau+\delta t_{j}-\delta t_{j-1}$ (for $j\geq1$ and
$t_{0}=\delta t_{0}\equiv0$). When all $\delta t_{j}\equiv0$, the repeated
process is reduced into equal time-interval measurements (ETIMs) formulated
in~\cite{purification}. After $N$ such ETIMs on the qubits and if all
measurement outcomes are $\left\vert g\right\rangle $, the density matrix of
the MR becomes $\rho_{m}^{(\tau)}(N)=V_{g}^{N}(\tau)\rho_{m}V_{g}^{\dagger
N}(\tau)/P_{g}^{(\tau)}(N)$, where $P_{g}^{(\tau)}(N)={\mathrm{Tr}}[V_{g}%
^{N}(\tau)\rho_{m}V_{g}^{\dagger N}(\tau)]$ is the survival
probability~\cite{purification,Wu-PRA}.
Here $V_{g}(\tau)\equiv\left\langle g\right\vert e^{-iH\tau/\hbar}\left\vert
g\right\rangle $ is an \emph{effective evolution operator} only acting on the MR.

$V_{g}(\tau)$ for model~(\ref{eq:3}) is diagonal in the basis \{$\left\vert
n\right\rangle $\}{:} $V_{g}(\tau)=\sum_{n\geqslant0}\lambda_{n}%
(\tau)\left\vert n\right\rangle \left\langle n\right\vert $. Here the
eigenvalues are $\lambda_{0}=e^{i{\Delta}\tau/{2} }$ and $\lambda
_{n}=e^{-i(n-1/2)\omega_{m}\tau}(\cos\Omega_{n}\tau+i\sin\Omega_{n}\tau
\cos2\theta_{n})$ with $\Omega_{n}=\sqrt{(\Delta-\omega_{m})^{2}/4+g^{2}n}$
for $n\geq1$. By carefully selecting the time interval $\tau$ such that
$\cos^{2}\Omega_{n}\tau\neq0$ (for $n\geq1$), all the values
\begin{equation}
|\lambda_{n}(\tau)|\ =\sqrt{1-\cos^{2}\Omega_{n}\tau\cos^{2}2\theta_{n}}%
\end{equation}
can be made less than $1$,
while $|\lambda_{0}(\tau)|$ always equals to $1$.

In the large $N$ limit for our specific JC model, the UTIMs result in
\begin{align*}
& V_{g}(\tau_{1})V_{g}(\tau_{2})...V_{g}(\tau_{N})=\sum_{n\geqslant0}%
|\bar{\lambda}_{n}|^{2N}\left\vert n\right\rangle \left\langle n\right\vert
\rightarrow\left\vert 0\right\rangle \left\langle 0\right\vert ,
\end{align*}
where $|\bar{\lambda}_{n}|^{2}=\left(  \left\vert \lambda_{n}(\tau
_{1})\right\vert ^{2}\left\vert \lambda_{n}(\tau_{2})\right\vert
^{2}...\left\vert \lambda_{n}(\tau_{N})\right\vert ^{2}\right)  ^{1/N}<1$ for
$n\geq1$ and $|\bar{\lambda}_{0}|=1$. All the density matrix elements of the
MR will vanish, except that of ground state $\left\vert 0\right\rangle $:
\begin{align}
\rho_{m}^{(\tau)}(N)  & =\sum_{n\geq0}|\bar{\lambda}_{n}|^{2N}\rho_{m}%
^{(n)}\left\vert n\right\rangle \left\langle n\right\vert /P_{g}^{(\tau)}(N)\\
& \rightarrow\left\vert 0\right\rangle \left\langle 0\right\vert ,~~\mathrm{
for}~N\rightarrow\infty,\nonumber
\end{align}
where $\rho_{m}^{(n)}\equiv\left\langle n\right\vert \rho_{m}\left\vert
n\right\rangle $ and the survival probability
$P_{g}^{(\tau)}(N)=\sum_{n\geqslant0}|\bar{\lambda}_{n}|^{2N}\rho_{m}%
^{(n)}\rightarrow\rho_{m}^{(0)}$. For the case of ETIMs, $|\bar{\lambda}%
_{n}|=\left\vert \lambda_{n}(\tau)\right\vert $. Repeated measurements drive
the evolution from the initial state (e.g., the thermal state) to the ground
state of a MR. The process has the same features as those of known ground
state cooling methods and is a new method for ground state cooling of a MR.
Furthermore, distinct from usual dynamical cooling schemes, the new cooling
method is based on repeated non-dynamical measurements.

\begin{figure}[th]
\includegraphics[bb=36 16 510 370,width=4.0cm]{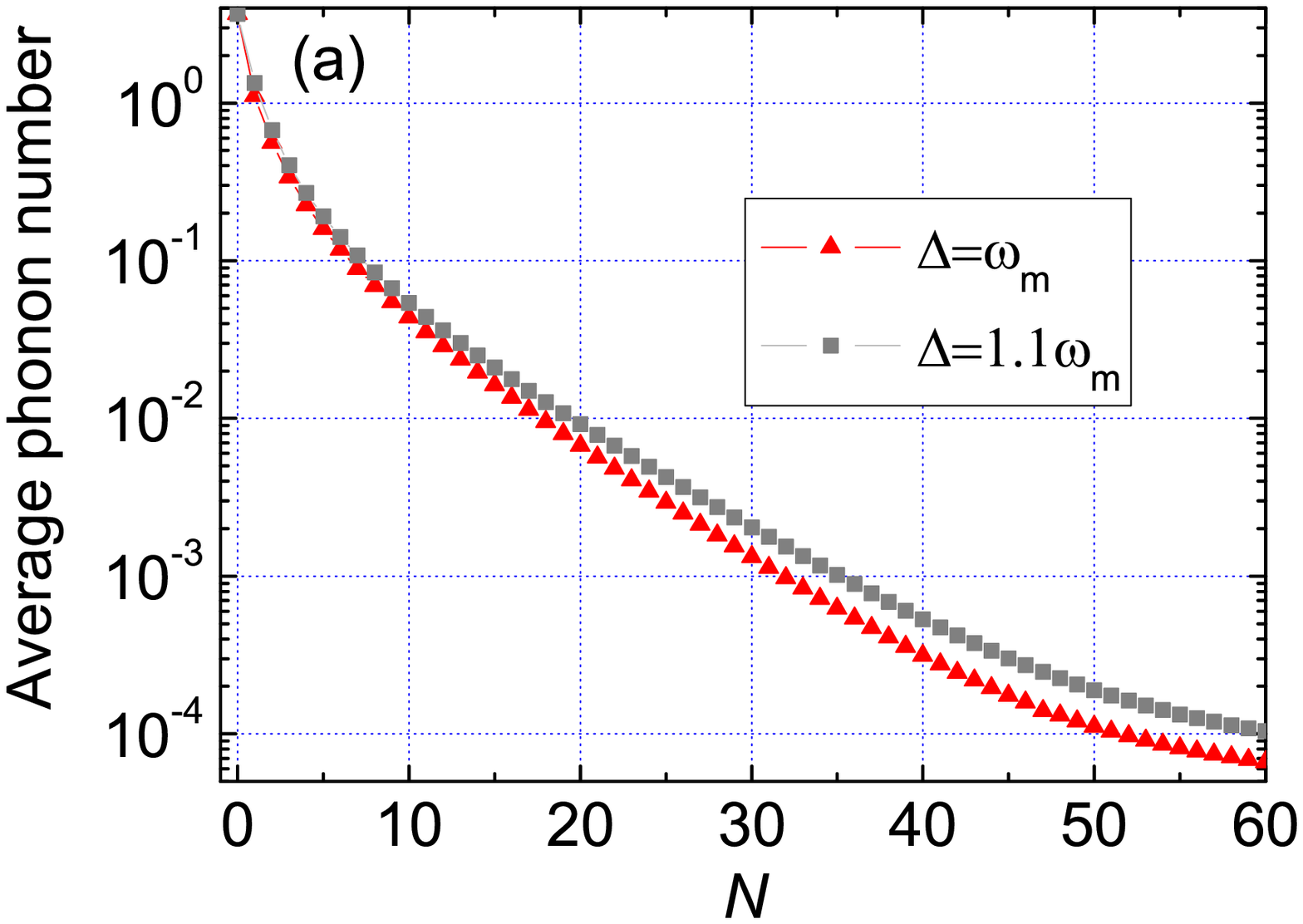}
\includegraphics[bb=36 16 510 370,width=4.0cm]{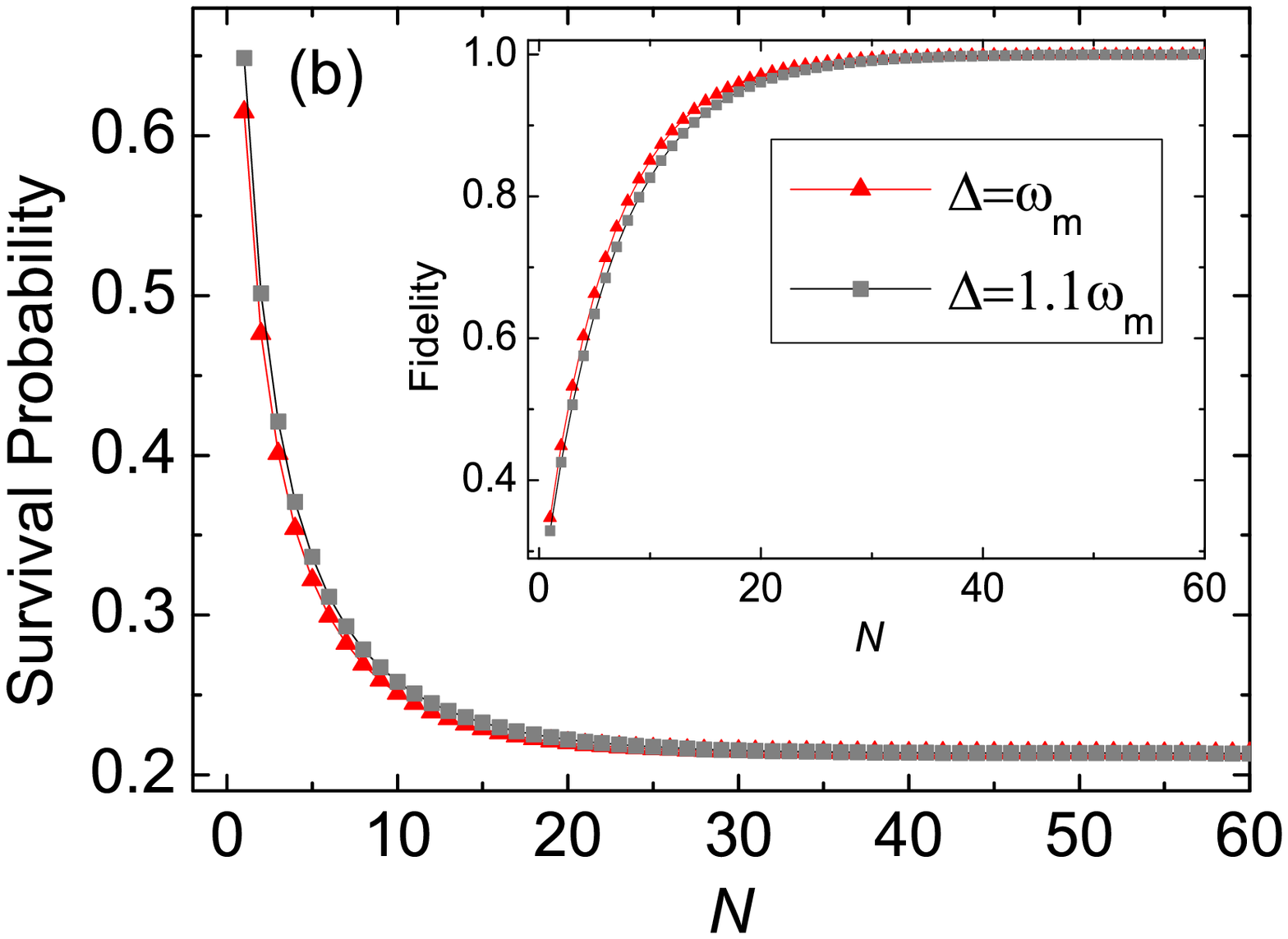}
\caption{(Color online) (a) The average phonon number $\bar{n}(N)$ after $N$
equal time-interval measurements (ETIMs) for the initial phonon number
$\bar{n}(0)=3.69$. (b) The corresponding survival probability $P_{g}^{(\tau
)}(N)$ and fidelity $F_{g}^{(\tau)}(N)$ (inset). The gray square lines denote
the resonant case $\Delta=\omega_{m}$. The red triangle lines denote the
non-resonant case $\Delta=1.1 \omega_{m}$. Here $g=0.04\omega_{m}$ and
$\tau=10/ \omega_{m}$. }%
\label{fig1}%
\end{figure}


\emph{Numerical results and robustness.}-- Consider a $2\pi\times100$ MHz
nano-mechanical resonator with quality factor $Q_{m}=10^{5}$ ($\gamma_{m}%
/2\pi=200$ Hz), coupled to a flux qubit with the tunneling splitting
$\Delta\simeq\omega_{m}$. The MR is initially at its thermal equilibrium state
at the ambient temperature $T=20$ mK, and the corresponding mean phonon number
is $\bar{n}(0)=1/[{{\exp(\hbar\omega_{m}/k_{B}T})-1}]=3.69$. Fig.~\ref{fig1}
shows for ETIMs the mean phonon number $\bar{n}(N)$, the survival probability
$P_{g}^{(\tau)}(N)$, and the fidelity $F_{g}^{(\tau)}(N)\equiv\left\langle
0\right\vert \rho_{m}^{(\tau)}(N)\left\vert 0\right\rangle $ as a function of
$N$, the number of measurements. The lines with red triangles in
Fig.~\ref{fig1} are for the on-resonant $\Delta=\omega_{m}$ case and the lines
with gray squares correspond to the off-resonant case, $\Delta=1.1\omega_{m}$.
The ground state cooling can be reached in both cases. Fig.~\ref{fig1} shows
that while the ground state cooling requires $60$ measurements ($\bar
{n}(N=60)\approx10^{-4}$), the mean phonon number decreases $90$ percent with
only five measurements.

\begin{figure}[th]
\includegraphics[bb=36 16 510 370,width=4.0cm]{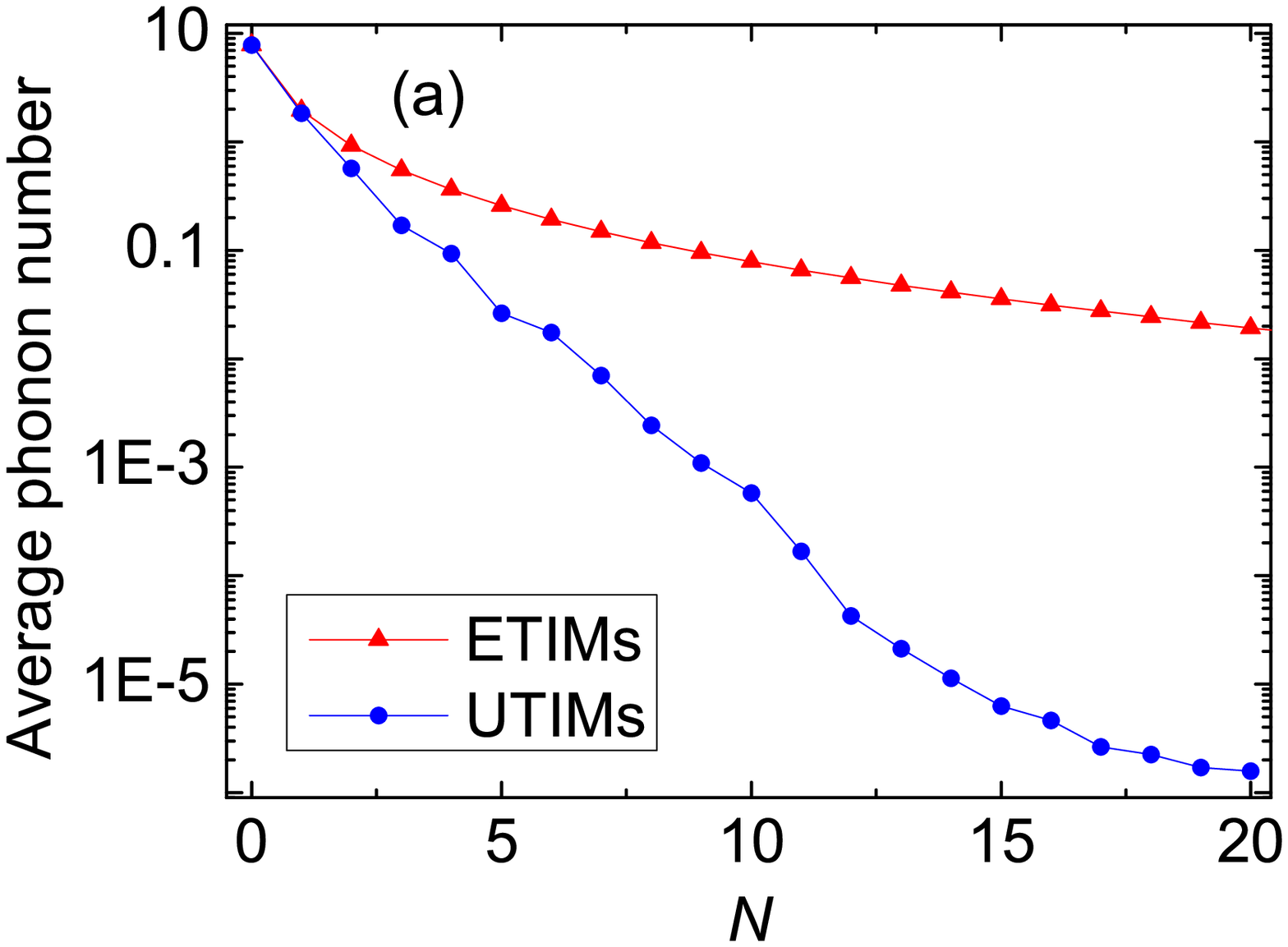}
\includegraphics[bb=36 16 510 370,width=4.0cm]{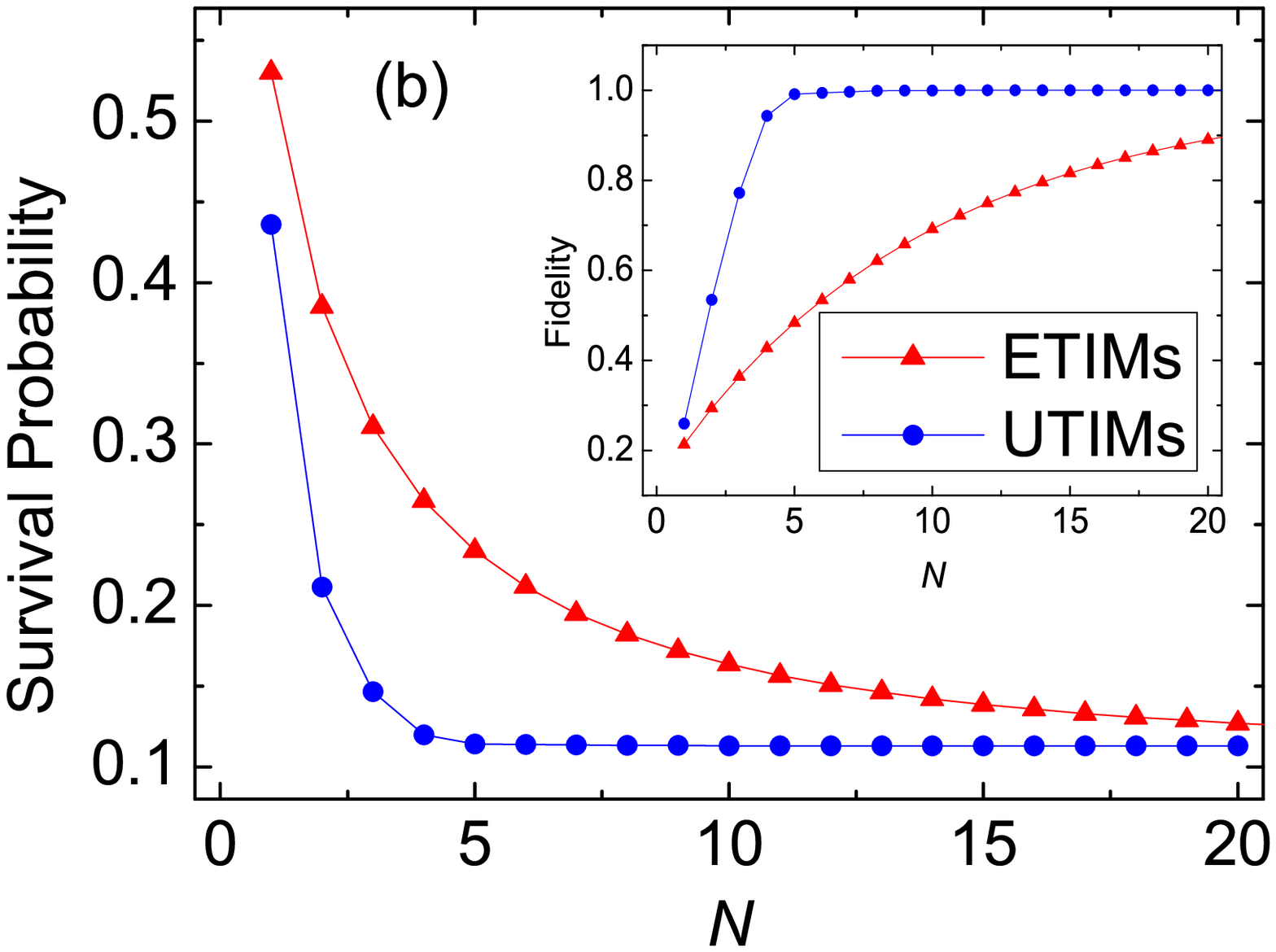}
\caption{(Color online) The average phonon number $\bar{n}(N)$ (a), the
survival probability (b), and the fidelity [inset in (b)], after $N$
measurements in the resonant case $\Delta=\omega_{m}$. The lines with red
triangles denote the equal time-interval measurements (ETIMs) ($\tau
=8/\omega_{m}$) and the lines with blue circles denote the unequal
time-interval measurements (UTIMs)~\cite{note1}. Here the initial phonon
number $\bar{n}(0)=7.84$, and the coupling strength $g=0.04\omega_{m}$.}%
\label{fig2}%
\end{figure}

The ideal ETIMs, with time interval $\tau$, may be difficult technically.
Fig.~\ref{fig2} shows, for UTIMs, the same physical quantities as
Fig.~\ref{fig1} for the $\Delta=\omega_{m}$ case, but with the higher bath
temperature $T=40$ mK and randomly selected time intervals $\tau_{j}$. The
mean phonon number is $\bar{n}(0)=7.84$ initially. It is noticeable that the
ground state cooling can be achieved much more efficiently than the ETIMs. It
is a remarkable advantage for experimentalists to achieve the ground state
cooling of MRs with random time intervals.

The physical reason is clear. In general, for a fixed $n$, the smaller the
maximal value $\Lambda_{n}=\max\{\left\vert \lambda_{n}(\tau_{j})\right\vert
\}_{j=1,...,N}$ is, the faster the term $|\bar{\lambda}_{n}|^{2N}\left\vert
n\right\rangle \left\langle n\right\vert $ decays with $N$. For ETIMs, it is
unavoidable that there exists $\Lambda_{n}$ ($n \geq1$) very close to one,
since for a given finite time interval $\tau$ the periodical function
$\cos(\Omega_{n}\tau)$ versus $\Omega_{n}$ runs across $0$ several times. The
corresponding component $\left\vert n\right\rangle \left\langle n\right\vert $
therefore decays very slowly. Specially, the component $\left\vert
n\right\rangle \left\langle n\right\vert $ will not decay with $N$ for ETIMs
when $\Lambda_{n}=1$. However for random UTIMs, the corresponding
$|\bar{\lambda}_{n}|^{2N}\left\vert n\right\rangle \left\langle n\right\vert $
decays faster since $|\bar{\lambda}_{n}|<\Lambda_{n}$ for any $n$.

Our cooling scheme is completely robust against the measurement operational
errors or randomness. It is also robust against the relaxation effect of the
MR. Our UTIMs can cool down a MR, initially at a thermal state with mean
phonon number $\lesssim10$ and a quality factor $Q_{m}=10^{5}$, in the time
interval $10\tau\approx100/\omega_{m}$, with $10$ measurements. It is $100$
times less than the MR's relaxation time, $\sim1/[\bar{n}(0)\gamma_{m}%
]\approx10000/\omega_{m}$. However, since the final survival probability is
proportional to the
initial population probability at the ground state, our UTIMs is more suitable
for further cooling based on a pre-cooled MR at a thermal (-like) state with a
smaller mean phonon number $\sim10$, e.g., by other cooling methods such as
the sideband cooling.

We should comment that there is an equal time-interval measurement-based
cooling of MR proposed in~\cite{Bergenfeldt-PRA}
using a Cooper pair box as the auxiliary, where the measurement effect is
averaged out and there is no explicit analytical expression for the process.
However, we find that the method fails to achieve the MR cooling for our model
after considerable number of equal time-interval measurements. We should also
remark that in the well-known sideband cooling the MR is coupled to a
high-frequency auxiliary with the faster damping rate. The energy flows from
the MR to the auxiliary and is then lost to the bath quickly. On the contrary,
the frequencies or the damping rates of the MR and the auxiliary qubit are of
the same order in our model. Our scheme is non-deterministic and based on
repeated non-dynamical measurements.

\emph{Implementation of the projective measurements on the flux qubit.}-- Our UTIMs
on flux qubit can be implemented by Josephson bifurcation amplifier (JBA) in a fast
and non-destructive way. As shown in Fig.~\ref{fig:circuit}, a JBA~\cite{Siddiqi2004}
(blue part), is coupled to the flux qubit inductively as the measurement device. The
JBA consists of a dc SQUID shunted by a capacitance $C$, subject to a microwave drive
$I_{RF} \cos(\omega_{d}t+\phi_{A})$. The JBA SQUID loop contains two Josephson
junctions of identical critical current $I_{A0}$ and different phase differences
$\varphi_{A1}$, $\varphi_{A2}$ respectively. The current in the loop is
$I_{A}=\bar{I}_{A}(\Phi_{A})\cos\varphi_{A}$, with $\Phi_{A}$, the flux bias in the
JBA SQUID, set by external coils, $\bar{I}_{A}(\Phi
_{A})=2I_{A0}\sin(\pi\Phi_{A}/\Phi_{0})$, and $\varphi_{A}=(\varphi
_{A1}+\varphi_{A2})/2$. The JBA circuit forms a driven resonator with nonlinear
Josephson inductance.

Since the JBA is positioned symmetrically with respect to the qubit loops $1$
and $2$, the two loops are coupled to the JBA with equal mutual inductance,
$M_{1}=M_{2}$. Due to the gradiometer design, the total qubit flux is
decoupled from the JBA~\cite{Paauw2009,Fedorov2010}. However, the JBA still
couples to the qubit loop $3$ through its influence on $\Phi_{3}$. If $\pi
M_{3}\langle{I}_{A}\rangle\ll\Phi_{0}$, this influence can be approximated as
a linear coupling to the $\tilde{\sigma}_{z}$ operator of the flux
qubit~\cite{Wang2010,Wang2011}, that is, an extra interaction term between the
qubit and the measurement device $H_{I}=\lambda(\Phi_{3b})\tilde{\sigma}%
_{z}\cos\varphi_{A}$.
Here the coupling coefficient is
$\lambda(\Phi_{3b})=-(\pi M_{3}\bar{I}_{A}/\Phi_{0})\kappa(\Phi_{3b})$
with
$\kappa(\Phi_{3b})=2\alpha_{0}\sin\left(  \pi(\Phi_{3b}/\Phi_{0})\right)
\left.  (d\Delta/d\alpha)\right\vert _{\alpha=\bar{\alpha}}$,
and $\bar{\alpha}=2\alpha_{0}\cos(\pi\Phi_{3b}/\Phi_{0})$; $\alpha_{0}$ is the ratio
between the Josephson energy of the smaller junctions and that of the two bigger
junctions in the flux qubit; $\Phi_{3b}$ is the total flux bias of the loop $3$. Thus
an external on-chip bias current $I_{B3}$ (green part in Fig.~\ref{fig:circuit}) can
be used to monitor the coupling strength $\lambda(\Phi_{3b})$.

Under a strong microwave drive, the Josephson energy of the junction
$-E_{JA}\cos\varphi_{A}$ is expanded beyond the harmonic approximation and the
classical dynamics can be described by a Duffing
oscillator~\cite{Landau_mechanics}. For a certain range of drive conditions,
the nonlinear oscillator exhibits bistable behavior with
hysteresis~\cite{Landau_mechanics,Siddiqi2004}. The two possible stable states
correspond to different oscillation amplitudes and phases, which can be
distinguished by transmitted or reflected microwave
signals~\cite{Boulant2007,Lupascu2007,Mallet2009}. Switching between the two
stable states happens when the driving power reaches a threshold. The
switching probability depends on the value of the nonlinear inductance, which
in our case depends on the states of the qubit. This is because the effective
Josephson energy of the junctions of the JBA is modified by the interaction
$H_{I}$
as $E_{JA}(\tilde{\sigma}_{z})=\bar{I}_{A}\Phi_{0}/2\pi-\lambda(\Phi
_{3b})\tilde{\sigma}_{z}$. Therefore, by measuring the phase of the
transmitted microwave signal, the state of the qubit is collapsed to one of
the eigenstates of its free Hamiltonian, $\tilde{\sigma}_{z}$ in our case.

In a realistic UTIMs, the measurement time, in the order of $\sim10$ ns,
should be considered. This is not a problem when measurements are performed as
follows: First, the qubit-MR interaction is switched off through the in-plane
magnetic field $B_{0}$; then a measurement pulse with readout and latching
plateau is sent to the JBA to readout the qubit state and induce the
projection to either $|g\rangle$ or $|e\rangle$ state; after this projection,
the qubit-MR interaction is switched on again. The process is repeated until
the MR reaches its ground state.

\emph{Conclusion.}-- We propose an ultrafast feasible scheme to cool a MR to
its ground state via repeated random-time-interval projective measurements on
an auxiliary flux qubit. The measurement scheme is almost independent of the
initial state of the MR.
It works when the MR couples with the qubit either on-resonance or
off-resonance, and even when the coupling $g(t)$ is time-dependent. The
cooling process is robust since it can be accomplished in a much shorter time
than the relaxation time.

Our scheme significantly simplifies experimental constraints since there is no
requirements for the control on the time intervals of measurements. In
principle, the MR can be cooled to arbitrarily low temperature with
arbitrarily small mean phonon number, within very short time.

\begin{acknowledgments}
We thank Dr. Hefeng Wang for helpful discussions. This work was supported by
the Research Funds of Renmin University of China (10XNL016), the Ikerbasque
Foundation Start-up, the Spanish MEC No. FIS2009-12773-C02-02, and the Basque
Government (grant IT472-10).
\end{acknowledgments}


\begin{thebibliography}{99}
\bibitem {note1}Here the time variances $\delta t_{N}$ ($N=1,...,20$) are
taken randomly, e.g., as \{0.26, 0.14, -0.40, -0.07, 0.06, 0.24, -0.20, -0.44,
-0.34, 0.01, -0.01, -0.12, 0.39, -0.28, -0.43, 0.06, -0.16, 0.17, 0.47, 0.01\}.
\end{thebibliography}
\end{document}